\ifdefined\WCLReview
\documentclass[12pt,draftclsnofoot,onecolumn]{IEEEtran}
\else
\documentclass[journal]{IEEEtran}
\fi
\usepackage{amsmath,amsfonts,amssymb}
\usepackage{algorithm}
\usepackage{algorithmic}
\usepackage{graphicx}
\usepackage{cite}
\usepackage{placeins}
\usepackage[colorlinks,citecolor=blue,linkcolor=blue]{hyperref}
\graphicspath{{figures/}}
\newcommand{\C}{\mathbb C}
\newcommand{\R}{\mathbb R}
\newcommand{\diag}{\operatorname{diag}}
\newcommand{\tr}{\operatorname{tr}}
\newcommand{\Proj}{\operatorname{\Pi}}

\begin{document}

\title{From Sparse Probes to Sum-Rate Maximization: Electromagnetic Twin Beamforming}

\author{Tuo~Wu, 
	Kangda Zhi,
        Jie~Tang,
        Jianchao Zheng, 
        Naofal Al-Dhahir, \IEEEmembership{Fellow,~IEEE,}
        and~Fumiyuki~Adachi,~\IEEEmembership{Life~Fellow,~IEEE}%
\thanks{Corresponding author: J. Tang.}
\thanks{T. Wu and J. Tang are with the School of Electronic and Information Engineering, South China University of Technology, Guangzhou 510640, China (E-mail: $ \rm  \{wutuo, eejtang\}@scut.edu.cn$). K. Zhi is with the School of Electrical Engineering and Computer Science, Technical University of Berlin, 10623 Berlin (E-mail: $\rm k.zhi@tu$-$\rm berlin.de$). J. Zheng is with the School of Computer Science and Engineering, Huizhou University, Huizhou 516007, China (E-mail: $\rm zhengjch@hzu.edu.cn$). Naofal Al-Dhahir is with the Department of Electrical and Computer Engineering, The University of Texas at Dallas, Richardson, TX 75080 USA (E-mail: $ \rm aldhahir@utdallas.edu$). F. Adachi is with the International Research Institute of Disaster Science (IRIDeS), Tohoku University, Sendai, Japan (E-mail: $\rm adachi@ecei.tohoku.ac.jp$). }}

\markboth{}{Wu \MakeLowercase{\textit{et al.}}: From Sparse Probes to Sum-Rate Maximization}
\maketitle

\begin{abstract}
\ifdefined\WCLReview\sloppy\fi
Probing every candidate location at each wireless-map update is costly. This letter develops electromagnetic-twin (ET) beamforming that converts sparse spatial probes into a sum-rate decision. The ET stores a location-dependent angular power spectrum, updates its innovation through graph-regularized estimation, and queries user covariances for projected beam optimization. A rate-sensitivity bound selects subsequent probes by query-weighted posterior-variance reduction. With 1\% random probes, ET beamforming achieves $2.69$ bit/s/Hz versus $1.12$ for a static channel knowledge map; with 7\% query-aware probes, it reaches $3.03$ bit/s/Hz, within 3.4\% of perfect-covariance beamforming.
\end{abstract}


\section{Introduction}
\IEEEPARstart{M}{ultiuser} beamforming depends on channel knowledge, but that knowledge does not evolve on a single time scale. Instantaneous channel state information (CSI) follows fast fading, whereas dominant arrival directions, spatial covariance, and blockage regions often change more slowly. The slower quantities can be learned across multiple sounding epochs and reused at locations where current measurements are sparse. Radio maps and channel knowledge maps (CKMs) exploit this separation by storing site-specific propagation information for wireless resource management~\cite{romeroCartography,biRadioMap,zengCKM,ckmTutorial}. Such environmental knowledge is increasingly relevant to sixth-generation (6G) systems~\cite{saad6G,wuFAS6G}.

Radio-map reconstruction infers unmeasured locations and exposes the completed map to communication algorithms~\cite{radioUNet,deepCompletion,radioGAT}. These methods remain useful when a representative current data set is available. In repeated operation, however, reacquiring such a data set over the complete service area can consume substantial sounding resources. Digital twins offer a complementary cycle: retain a state, update it from new physical evidence, query it for a decision, and synchronize it again~\cite{nguyenTwin,khanTwin,alkhateebTwin,wangDTC}. A CKM describes the wireless knowledge being stored; an electromagnetic twin (ET) adds the persistent update--query--measurement loop. Thus, the ET is not proposed as a replacement for CKM reconstruction, but as a way to operate and refresh that knowledge over time.

Persistence alone does not determine where to measure. We call each discrete spatial bin in the service-area grid a \emph{cell}; a cell is a candidate probe location, not a cellular-network base station. Under a small budget, a probe far from active users may reduce global map error but contribute little to their beams, whereas a probe near a propagation change may improve the relevant covariances substantially. Therefore, minimizing a uniform reconstruction error and maximizing a communication utility are not equivalent measurement objectives. We ask: \emph{which cells should be probed so that the updated ET supports the largest next-round sum rate?}

To answer this question, we store a location-dependent angular power spectrum (APS). A scalar gain map does not retain array directions, whereas a map of instantaneous complex channels changes too rapidly to be persistent. The APS provides a middle layer: selected uplink probes update its innovation, and a linear array query converts the updated APS into user channel covariances. These covariances enter power-constrained sum-rate maximization, while query-weighted update uncertainty selects the next probes. The ET therefore complements fast CSI with persistent second-order knowledge rather than replacing instantaneous channel tracking. Our contributions are summarized as follows.
\begin{itemize}
\item We formulate a two-stage next-update sum-rate problem: probes are selected before their observations, and beams are optimized after those observations update the ET.
\item A graph-regularized estimator returns an APS and its uncertainty; APS queries return user covariances; and projected successive convex approximation (SCA) maximizes the covariance-based sum rate.
\item A sum-rate sensitivity bound yields a rate-relevant probe rule. A nested Monte Carlo evaluation separates noisy map updates from fast fading and isolates memory, density, and query-aware selection.
\end{itemize}

\section{Electromagnetic-Twin System Model}
\begin{figure*}[t]
\centering
\includegraphics[width=0.68\textwidth]{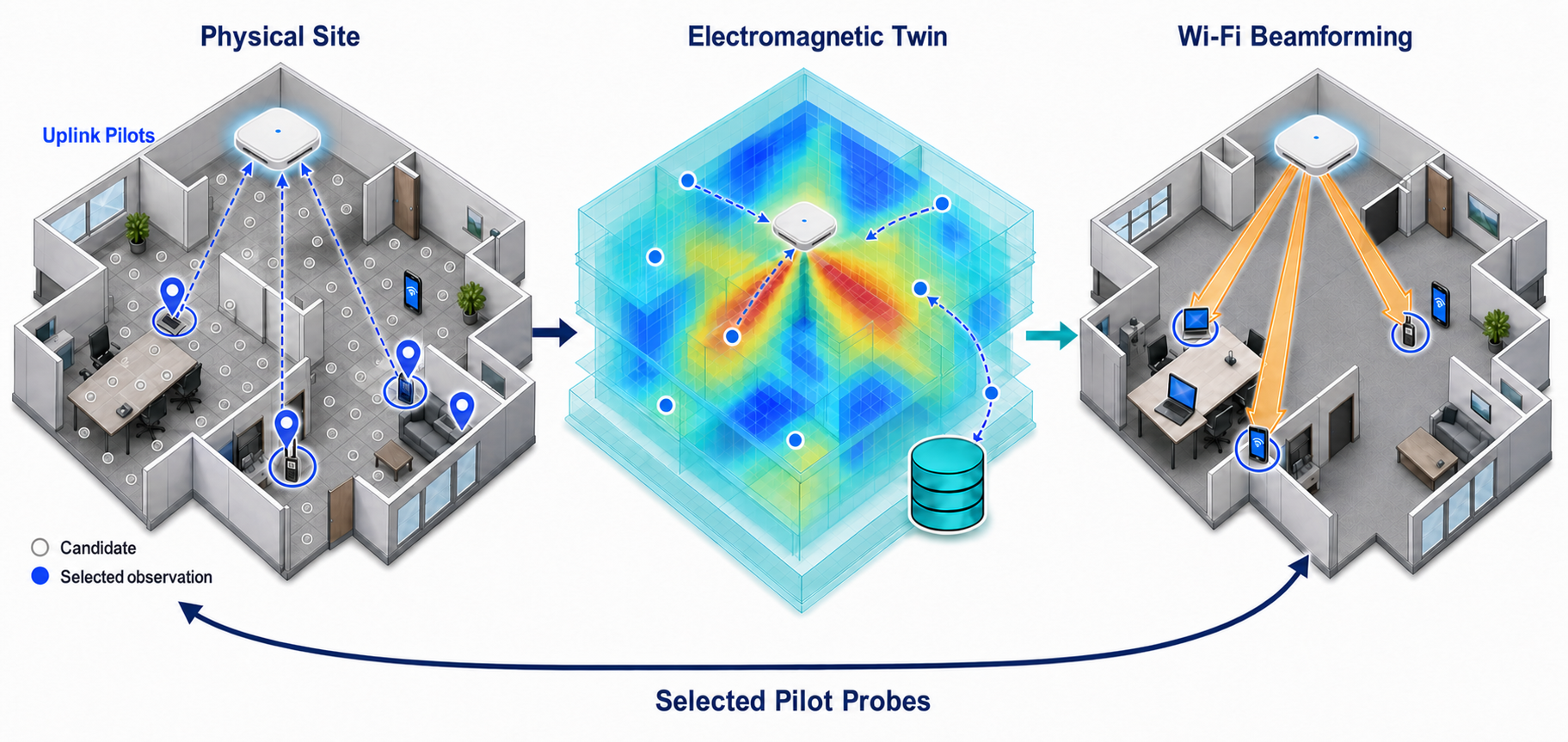}
\caption{From sparse probes to ET beamforming in an indoor Wi-Fi site. Gray circles are candidate probe cells, blue pins are selected cells, and the users queried for downlink beamforming are shown in the right panel. AI-generated schematic; see Acknowledgment.}
\label{fig:loop}
\end{figure*}

Fig.~\ref{fig:loop} shows the chain in an indoor Wi-Fi site. At update epoch $t$, one $M$-antenna access point (AP) serves $K$ single-antenna users over $N$ candidate probe cells. The set $\Omega_t\subseteq\{1,\ldots,N\}$ contains the blue cells selected for sounding; gray cells are not measured. Probes at $\Omega_t$ transmit mutually orthogonal uplink pilots of length at least $|\Omega_t|$, allowing the AP to separate their directional-power observations. The observations update the stored ET, queries at the fixed user locations provide downlink covariances, and posterior uncertainty selects the next probe batch. In contrast, a static CKM reconstructs a map from current samples but neither retains the previous state nor closes this update--query--probe loop.

\begin{table}[t]
\caption{Operational Difference Between Static CKM and ET}
\label{tab:ckm-et}
\centering
\scriptsize
\setlength{\tabcolsep}{2.5pt}
\begin{tabular}{lcc}
\hline
 & Static CKM & Proposed ET \\
\hline
Input & current probes & stored state + current probes \\
Estimate & complete current map & state innovation \\
Output & reconstructed map & APS, covariances, next probes \\
Temporal role & one-shot reconstruction & persistent update--query loop \\
\hline
\end{tabular}
\end{table}

\subsection{Persistent Propagation State}
Divide the service area into $N$ candidate probe cells and the angular domain into $L$ directions $\{\vartheta_\ell\}_{\ell=1}^{L}$. The entry $S_t(n,\ell)$ represents the average received power at cell $n$ associated with direction $\vartheta_\ell$. Collecting these entries gives the persistent ET state
\begin{equation}
 \mathbf S_t=\big[S_t(n,\ell)\big]_{n,\ell}\in\R_+^{N\times L},
 \label{eq:state}
\end{equation}
where row $\mathbf S_t(n,:)$ is the APS at cell $n$. Unlike a scalar power map, it preserves the directional information needed by an antenna array; unlike instantaneous complex CSI, it is suitable for storage across update intervals. Sparse observations update $\widehat{\mathbf S}_{t-1}$ to $\widehat{\mathbf S}_t$, state queries determine the beam matrix $\mathbf W_t$, and posterior uncertainty determines the next probe set.

Let $\mathbf P_t\in\{0,1\}^{|\Omega_t|\times N}$ select the APS rows indexed by $\Omega_t$. Each selected probe contributes one $L$-dimensional directional-power vector. After orthogonal-pilot separation and directional processing at the AP, the collected observations satisfy
\begin{equation}
 \mathbf Y_t=\mathbf P_t\mathbf S_t+\mathbf N_t,
 \label{eq:measure}
\end{equation}
where $\mathbf Y_t,\mathbf N_t\in\R^{|\Omega_t|\times L}$ and $\mathbf N_t$ represents finite-pilot and directional-power extraction errors. Because the ET retains $\widehat{\mathbf S}_{t-1}$, the sparse current probes estimate only the innovation
\begin{equation}
 \boldsymbol\Delta_t=\mathbf S_t-\widehat{\mathbf S}_{t-1},
 \label{eq:innovation}
\end{equation}
Subtracting the stored prediction from the new probes gives
\begin{equation}
 \mathbf E_t=\mathbf Y_t-\mathbf P_t\widehat{\mathbf S}_{t-1}
 =\mathbf P_t\boldsymbol\Delta_t+\mathbf N_t.
 \label{eq:residual}
\end{equation}
The desired innovation is zero in unchanged regions, so previously learned knowledge is retained rather than reconstructed from a few current samples. A graph Laplacian $\mathbf L\succeq\mathbf0$ represents correlation between neighboring cells~\cite{shumanGSP} and allows a detected local change to influence nearby unobserved cells.

\subsection{APS-to-Covariance Query}
The AP serves the $K$ users at cells $n_1,\ldots,n_K$. With normalized array response $\mathbf a(\vartheta_\ell)\in\C^M$ and dictionary $\mathbf A=[\mathbf a(\vartheta_1),\ldots,\mathbf a(\vartheta_L)]\in\C^{M\times L}$, the APS powers weight the directional array responses. Querying the ET at user $k$ therefore gives
\begin{align}
 \mathbf B_{k,t}&=\mathbf A\diag(\widehat{\mathbf S}_t(n_k,:))\mathbf A^H,\nonumber\\
 \widehat{\mathbf R}_{k,t}&=\mathbf B_{k,t}
 +\epsilon_R\frac{\tr(\mathbf B_{k,t})}{M}\mathbf I_M,
\label{eq:covariance}
\end{align}
where $\mathbf B_{k,t},\widehat{\mathbf R}_{k,t}\in\C^{M\times M}$. The first term superposes the directional covariance components. The second term is isotropic diagonal loading: it represents unresolved diffuse power and prevents a rank-deficient covariance from making the beam design numerically fragile; the trace is evaluated before loading. The current channel follows $\mathbf h_{k,t}\sim\mathcal{CN}(\mathbf0,\mathbf R_{k,t})$. Hence, the ET converts sparse physical observations into beamforming knowledge through $\mathbf Y_t\rightarrow\widehat{\mathbf S}_t\rightarrow\widehat{\mathbf R}_{k,t}$.

At epoch $t$, let $x_{k,t}$ be a unit-power data symbol and let $\mathbf W_t=[\mathbf w_{1,t},\ldots,\mathbf w_{K,t}]\in\C^{M\times K}$ be the downlink beam matrix. User $k$ receives
\begin{equation}
 y_{k,t}=\mathbf h_{k,t}^H\mathbf w_{k,t} x_{k,t}+
 \sum_{j\ne k}\mathbf h_{k,t}^H\mathbf w_{j,t}x_{j,t}+z_{k,t},
 \label{eq:received}
\end{equation}
where $z_{k,t}\sim\mathcal{CN}(0,\sigma^2)$ is receiver noise. Using the queried covariance as second-order channel knowledge gives the moment-based rate
\begin{equation}
 \bar r_{k,t}(\mathbf W_t)=\log_2\!\left(1+
 \frac{\mathbf w_{k,t}^H\widehat{\mathbf R}_{k,t}\mathbf w_{k,t}}
 {\sigma^2+\sum_{j\ne k}\mathbf w_{j,t}^H\widehat{\mathbf R}_{k,t}\mathbf w_{j,t}}\right).
 \label{eq:moment-rate}
\end{equation}
The operational objective is the sum rate rather than map normalized mean-square error (NMSE). For the state obtained from the current sparse probes, the AP solves
\begin{equation}
 \begin{aligned}
 (\mathrm P_{\rm BF})\quad
 \max_{\mathbf W_t}\quad &F_t(\mathbf W_t;\widehat{\mathbf S}_t)
 =\sum_{k=1}^{K}\bar r_{k,t}(\mathbf W_t)\\
 \mathrm{s.t.}\quad &\|\mathbf W_t\|_F^2\le P,
 \end{aligned}
 \label{eq:beam-problem}
\end{equation}
Here, $P$ is the AP's total downlink transmit-power budget. Through~\eqref{eq:covariance}, $\widehat{\mathbf S}_t$ determines every $\widehat{\mathbf R}_{k,t}$. Thus, different probe observations produce different covariance estimates, beams, and achievable sum rates.

The remaining question is which $B$ cells should be probed next. Let $\mathcal B_t$ denote this new batch and let $\mathcal U_t(\mathcal B_t)\triangleq\mathcal U(\widehat{\mathbf S}_t,\mathcal B_t,\mathbf Y_{t+1})$ denote the state produced from the stored ET and its future observations. Assuming that the queried users remain active over the next update interval, the ideal finite-budget problem is
\begin{align}
 (\mathrm P_{\rm loop})\quad
 \max_{\mathcal B_t}
 &\ \mathbb E_{\mathbf Y_{t+1}}
 \!\left[F_{t+1}(\mathbf W_{t+1}^{\star};\widehat{\mathbf S}_{t+1})\right]
 \nonumber\\
 \mathrm{s.t.}\quad
 &\mathcal B_t\subseteq\{1,\ldots,N\}\setminus\Omega_t,\nonumber\\
 &|\mathcal B_t|\le B,\nonumber\\
 &\widehat{\mathbf S}_{t+1}=\mathcal U_t(\mathcal B_t),\nonumber\\
 &\mathbf W_{t+1}^{\star}\in
 \arg\max_{\|\mathbf W\|_F^2\le P}
 F_{t+1}(\mathbf W;\widehat{\mathbf S}_{t+1}).
 \label{eq:loop-problem}
\end{align}
Problem~$(\mathrm P_{\rm loop})$ is a two-stage stochastic program. The first-stage decision $\mathcal B_t$ is made before $\mathbf Y_{t+1}$ is observed. After those measurements update the ET, the second-stage recourse $\mathbf W_{t+1}^{\star}$ maximizes the resulting sum rate. Problem~$(\mathrm P_{\rm BF})$ exposes this beamforming stage. Since probe selection is discrete and future observations are unknown, we solve the update and beam layers directly and derive a deterministic rate-relevant surrogate for the first stage. This causal order prevents the estimator from using future observations noncausally to inflate the reported rate.

\section{From Sparse Probes to Sum-Rate Maximization}
\subsection{Updating the ET from Sparse Probes}
We first instantiate $\mathcal U_t$ in~$(\mathrm P_{\rm loop})$. Because the previous APS is retained, the current observations estimate the innovation through the graph-regularized convex quadratic problem~\cite{shumanGSP}
\begin{align}
 \widehat{\boldsymbol\Delta}_t=\arg\min_{\boldsymbol\Delta\in\R^{N\times L}}
 &\ \frac{\mu}{2}\|\mathbf P_t\boldsymbol\Delta-\mathbf E_t\|_F^2
 +\frac{\lambda}{2}\tr(\boldsymbol\Delta^T\mathbf L\boldsymbol\Delta) \nonumber\\
 &+\frac{\eta+\varepsilon}{2}\|\boldsymbol\Delta\|_F^2,
 \label{eq:update-problem}
\end{align}
where $\mu>0$ is the assumed observation-noise precision (inverse normalized variance), $\lambda$ propagates an observed change over neighboring cells, $\eta$ shrinks unsupported innovations toward zero and thereby favors persistence, and $\varepsilon>0$ ensures strict positive definiteness. Thus, the first term fits the selected probes, the second enforces graph smoothness, and the third prevents sparse evidence from changing the entire stored map. The normal equations yield
\begin{align}
 \mathbf H_t&=\mu\mathbf P_t^T\mathbf P_t+\lambda\mathbf L
 +(\eta+\varepsilon)\mathbf I_N,\label{eq:hessian}\\
 \widehat{\boldsymbol\Delta}_t&=\mathbf H_t^{-1}\mu\mathbf P_t^T\mathbf E_t,\label{eq:closed-update}\\
 \widehat{\mathbf S}_t&=[\widehat{\mathbf S}_{t-1}+\widehat{\boldsymbol\Delta}_t]_+.
 \label{eq:state-update}
\end{align}
The update also quantifies uncertainty for future probing. Let $\boldsymbol\delta_\ell$ and $\mathbf e_\ell$ be column $\ell$ of $\boldsymbol\Delta_t$ and $\mathbf E_t$, and define $\boldsymbol\Lambda=\lambda\mathbf L+(\eta+\varepsilon)\mathbf I_N$. Under the Gaussian working model with precision $\mu$ and prior $\boldsymbol\delta_\ell\sim\mathcal N(\mathbf0,\boldsymbol\Lambda^{-1})$,
\begin{equation}
 p(\boldsymbol\delta_\ell\mid\mathbf e_\ell)
 \propto\exp\!\left[-\frac{1}{2}
 \left(\mu\|\mathbf P_t\boldsymbol\delta_\ell-\mathbf e_\ell\|_2^2
 +\boldsymbol\delta_\ell^T\boldsymbol\Lambda\boldsymbol\delta_\ell\right)\right].
 \label{eq:posterior}
\end{equation}
Thus,~\eqref{eq:closed-update} is the posterior mean and maximum a posteriori (MAP) estimate, and $\mathbf V_t=\mathbf H_t^{-1}$ is the covariance shared by all angular bins. Clipping enforces nonnegative power, while one factorization updates all $L$ bins. This stage returns both the APS for beamforming and uncertainty for probing. The static CKM instead estimates $\mathbf S_t$ from $(\mathbf P_t,\mathbf Y_t)$ without memory.

\subsection{Sum-Rate Maximization with ET Queries}
After sparse probes fix $\widehat{\mathbf S}_t$ and its queried covariances,~$(\mathrm P_{\rm BF})$ remains nonconvex but admits a projected SCA update. Within this subsection, the epoch index $t$ is suppressed for readability. Define
\begin{align}
 T_k&=\sigma^2+\sum_{j=1}^{K}\mathbf w_j^H\widehat{\mathbf R}_k\mathbf w_j,\\
 I_k&=T_k-\mathbf w_k^H\widehat{\mathbf R}_k\mathbf w_k.
\end{align}
Here, $T_k$ is user $k$'s total received signal-plus-interference-and-noise power, whereas $I_k$ excludes the desired-stream power and is therefore its interference-plus-noise power. Since $\bar r_k=\log_2T_k-\log_2I_k$, differentiating these two terms gives the Wirtinger gradient
\begin{equation}
 \mathbf g_j=\frac{1}{\ln2}\sum_{k=1}^{K}\widehat{\mathbf R}_k\mathbf w_j
 \left(\frac{1}{T_k}-\frac{\mathbf{1}_{\{j\ne k\}}}{I_k}\right).
 \label{eq:gradient}
\end{equation}
Stack the user gradients as $\mathbf G=[\mathbf g_1,\ldots,\mathbf g_K]$. At iteration $q$, use the proximal minorizer
\begin{align}
 \widetilde F_q(\mathbf W)=F(\mathbf W^{(q)})
 &+2\Re\!\left\{\tr\!\left[(\mathbf G^{(q)})^H
 (\mathbf W-\mathbf W^{(q)})\right]\right\}\nonumber\\
 &-\tau_q\|\mathbf W-\mathbf W^{(q)}\|_F^2,
 \label{eq:beam-minorizer}
\end{align}
where backtracking increases $\tau_q>0$ until the candidate satisfies $F(\mathbf W)\ge\widetilde F_q(\mathbf W)$. Maximizing this concave quadratic over the transmit-power ball is a convex subproblem. Completing the square gives
\begin{equation}
 \mathbf W^{(q+1)}=\Proj_{\|\mathbf W\|_F^2\le P}
 \left(\mathbf W^{(q)}+\frac{1}{\tau_q}\mathbf G^{(q)}\right),
 \label{eq:projected-step}
\end{equation}
The projection rescales $\mathbf W$ only if its power exceeds $P$. Minorizer tightness and backtracking make the sum rate nondecreasing, with stationary limit points under standard majorization--minimization (MM)/SCA conditions~\cite{scutariSCA}. Generalized-eigenvector beams initialize the method. Weighted minimum mean-square error (WMMSE) is another option, but the covariance-moment rate requires a tailored auxiliary-variable reformulation~\cite{christensenWMMSE}; projected SCA exposes the ET covariance directly.

\subsection{Selecting Rate-Relevant Probes}
For the discrete outer decision $\mathcal B_t$, a probe is useful when its uncertainty is large and relevant to an active user. Define
\begin{align}
 q_i&=\max_k\exp\!\left(-\frac{\|\mathbf x_i-\mathbf u_k\|_2^2}{2r_q^2}\right),\nonumber\\
 \mathbf Q_t&=\diag(1+\rho q_1,\ldots,1+\rho q_N).
 \label{eq:query-weight}
\end{align}
where $\mathbf x_i$ and $\mathbf u_k$ are the coordinates of candidate probe cell $i$ and user $k$, respectively. Since~\eqref{eq:covariance} is linear in the APS and $\sigma^2>0$, the sum rate is locally Lipschitz over the compact power set. For finite $c_t$,
\begin{equation}
 \mathbb E|F_t(\mathbf W;\mathbf S_t)-F_t(\mathbf W;\widehat{\mathbf S}_t)|^2
 \le c_tL\tr(\mathbf Q_t\mathbf V_t).
 \label{eq:rate-bound}
\end{equation}
Thus, query-weighted uncertainty bounds communication-objective error, giving the surrogate
\begin{equation}
 (\mathrm P_{\rm VR})\quad
 \min_{\substack{\mathcal B_t\cap\Omega_t=\varnothing\\|\mathcal B_t|\le B}}
 \tr\!\left(\mathbf Q_t\mathbf V_t(\mathcal B_t)\right),
 \label{eq:variance-problem}
\end{equation}
where $\mathbf V_t(\mathcal B_t)$ is the innovation covariance after adding batch $\mathcal B_t$. For a candidate cell $i$, the matrix inversion lemma gives the updated covariance and its exact one-probe reduction in~$(\mathrm P_{\rm VR})$:
\begin{align}
 \mathbf V_i^+&=\mathbf V-
 \frac{\mu\mathbf V\mathbf e_i\mathbf e_i^T\mathbf V}
 {1+\mu\mathbf e_i^T\mathbf V\mathbf e_i},\label{eq:sherman}\\
 a_i&=\tr[\mathbf Q_t(\mathbf V-\mathbf V_i^+)]
 =\frac{\mu\mathbf e_i^T\mathbf V\mathbf Q_t\mathbf V\mathbf e_i}
 {1+\mu\mathbf e_i^T\mathbf V\mathbf e_i}.
 \label{eq:active-score}
\end{align}
Select $i^\star=\arg\max_i a_i$, set $\mathbf V\leftarrow\mathbf V_{i^\star}^+$, and repeat $B$ times. With diagonal covariance this reduces to query-weighted maximum variance; otherwise,~\eqref{eq:active-score} includes uncertainty reduction at correlated cells. Equation~\eqref{eq:rate-bound} therefore ties each probe to the next sum-rate decision.



\section{Numerical Results}
\subsection{Setup and Baselines}
We use a normalized narrowband indoor model on a $20\times15$ grid with 1-m cells. At 5.2 GHz, a half-wavelength-spaced $M=8$ element ULA is fixed at AP coordinate $(0,8)$ outside the left boundary. The $K=4$ users remain at $(7,6)$, $(10,12)$, $(14,7)$, and $(17,10)$ for every method, and $L=15$ angular bins span $[-70^\circ,70^\circ]$. Each APS combines direct, reflected, and diffuse components, so propagation is not obstacle-free LoS. Between epochs, the region $10\le x\le17$, $5\le y\le11$ attenuates its direct component and gains a shifted reflection; the last two users lie in this region. User cells are excluded from probe candidates to prevent trivial covariance observation.

The parameters are $\mu=24$, $(\lambda,\eta)=(0.8,0.2)$, static-CKM weight $\lambda_s=0.72$, $(\rho,r_q)=(50,1.5)$, $\epsilon_R=0.01$, and 50 beam iterations. Unless stated otherwise, normalized downlink SNR $P/\sigma^2=10$ dB; large-scale path loss is absorbed into the APS covariance scale, so this is an effective received-domain SNR rather than raw AP power. APS observations contain 10\% signal-dependent Gaussian error plus a 2\% floor to model finite-pilot and power-extraction errors. Each point uses 20 independent noisy APS updates and 25 fast-fading draws per update: 500 common Monte Carlo realizations, not a system parameter, shared by all methods.

\emph{Perfect covariance} uses true user covariances. The \emph{ET} updates its APS by~\eqref{eq:update-problem}; the \emph{static CKM} uses the same probes without memory; and the \emph{stale twin} ignores new probes. A common beam solver isolates the value of propagation knowledge.

\subsection{Sparse Probes and Persistent Wireless Knowledge}
\begin{figure}[t]
\centering
\includegraphics[width=0.62\columnwidth]{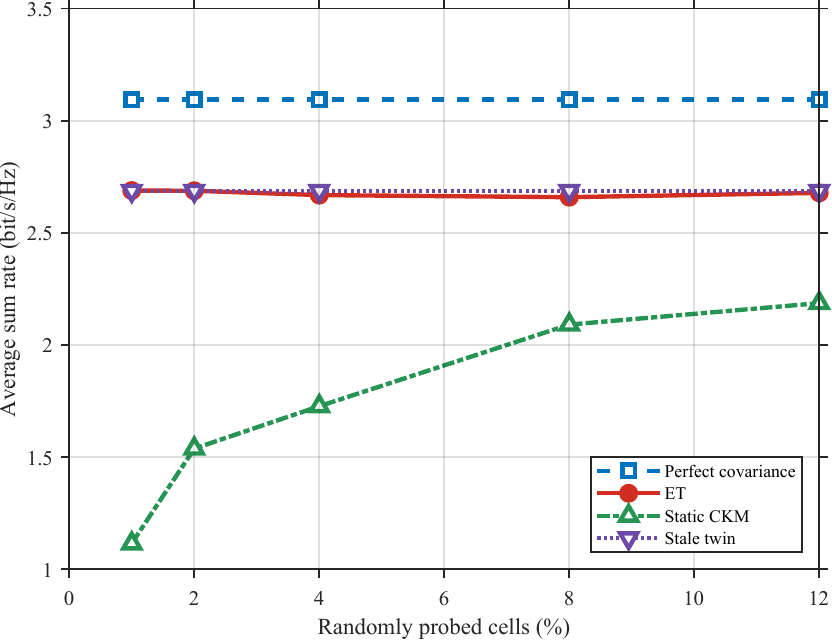}
\caption{500-realization average sum rate versus random probe percentage.}
\label{fig:pilot-rate}
\end{figure}

Fig.~\ref{fig:pilot-rate} uses random probing to isolate ET persistence. At 1\%, ET beamforming achieves $2.69$ bit/s/Hz versus $1.12$ bit/s/Hz for static CKM, for which sparse current observations cannot reconstruct the complete field. ET remains close to the stale twin because uniform probes may reduce site-wide error without observing the changed users' neighborhoods and improving their covariances. This motivates the query-aware probing in Fig.~\ref{fig:loop-rate}.

\begin{figure}[t]
\centering
\includegraphics[width=0.62\columnwidth]{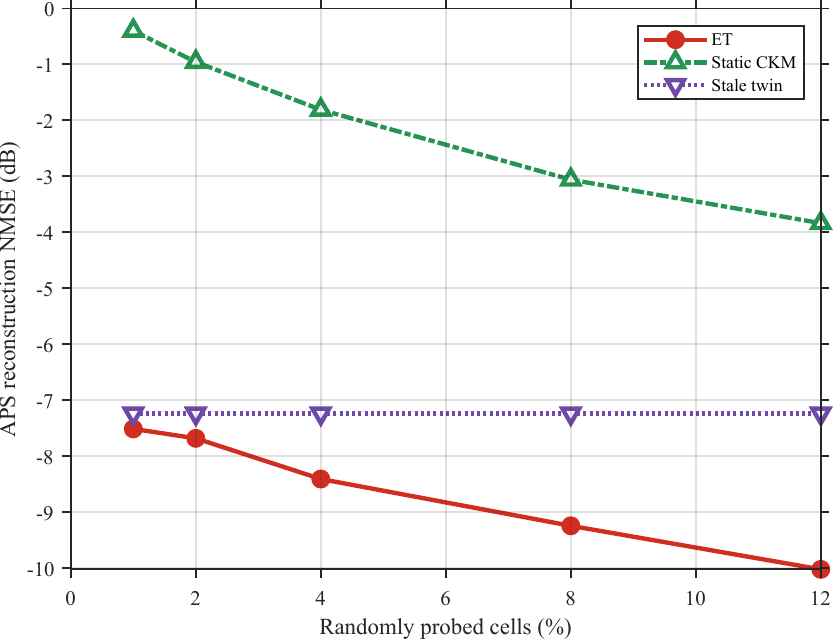}
\caption{20-update average APS NMSE versus random probe percentage.}
\label{fig:pilot-nmse}
\end{figure}

APS NMSE is $10\log_{10}(\|\widehat{\mathbf S}_t-\mathbf S_t\|_F^2/\|\mathbf S_t\|_F^2)$. In Fig.~\ref{fig:pilot-nmse}, ET improves from $-7.51$ dB at 1\% to $-10.02$ dB at 12\%, whereas the stale twin remains at $-7.24$ dB. Static CKM improves with dense current evidence; ET's advantage is combining sparse new evidence with stored knowledge.

\subsection{Sum-Rate Benefit of ET Covariances}
\begin{figure}[t]
\centering
\includegraphics[width=0.85\columnwidth]{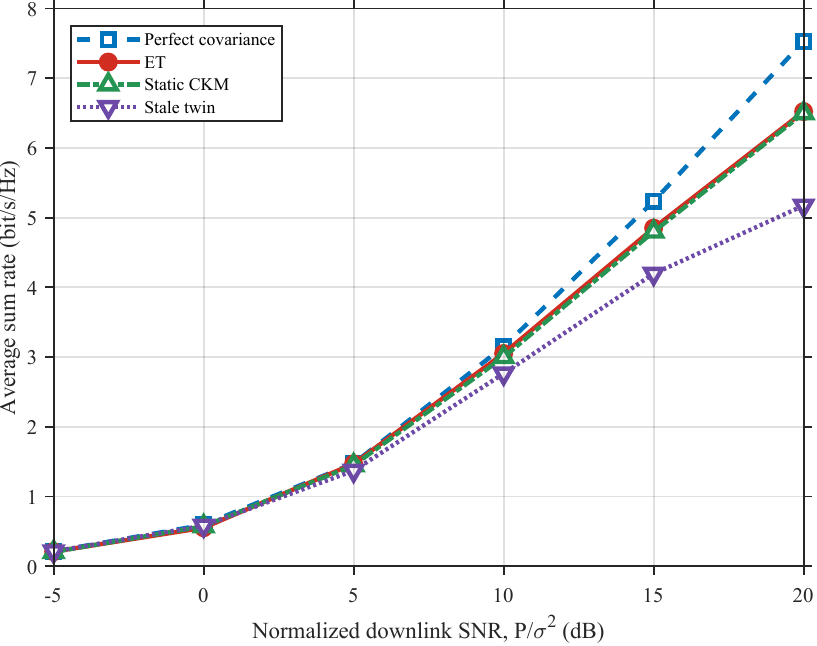}
\caption{500-realization average sum rate versus normalized SNR after 7\% query-aware probing.}
\label{fig:snr}
\end{figure}

Fig.~\ref{fig:snr} fixes the same four users and 7\% query-aware probes for every method while normalized SNR varies from $-5$ to 20 dB. Covariance errors matter little in the noise-limited regime but create multiuser leakage at high SNR. At 20 dB, ET attains $6.52$ bit/s/Hz versus $5.17$ bit/s/Hz for the stale twin, a 26.1\% gain. Static CKM gives a similar queried-user sum rate because probes concentrate near those users, although Fig.~\ref{fig:pilot-nmse} shows worse site-wide APS accuracy.

\subsection{Sum-Rate-Driven Probe Selection}
\begin{figure}[t]
\centering
\includegraphics[width=0.850\columnwidth]{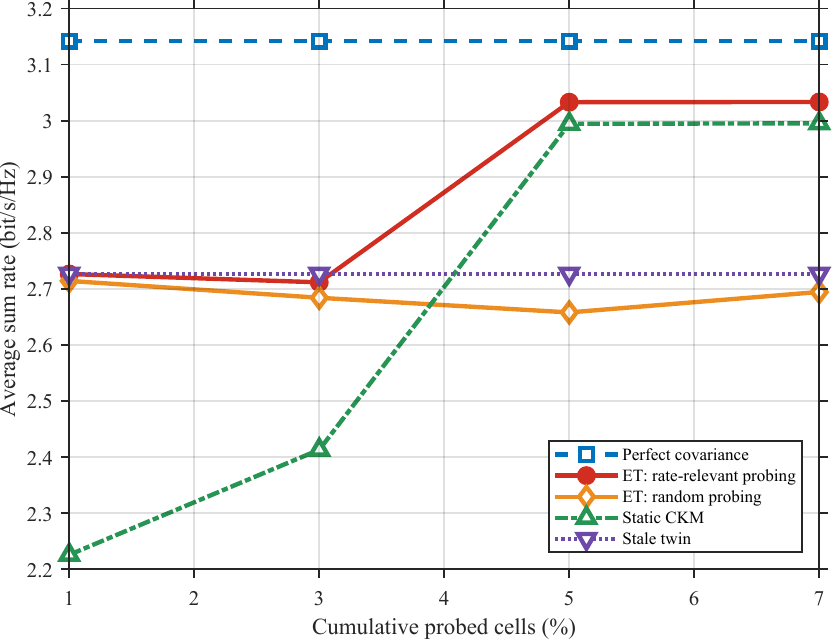}
\caption{500-realization average closed-loop sum rate under rate-relevant and random probing.}
\label{fig:loop-rate}
\end{figure}

The closed loop starts from 1\% observed cells and adds 2\% per update. Each round appends new probes to earlier observations, recomputes the innovation and user covariances, and redesigns the beams; no weights are trained online. At 7\% in Fig.~\ref{fig:loop-rate}, rate-relevant probing gives $3.03$ bit/s/Hz, within 3.4\% of perfect covariance, versus $2.70$ for random probing. Static CKM reaches $3.00$ because probes are query focused, but this local utility does not imply accurate full-field reconstruction.

\begin{figure}[t]
\centering
\includegraphics[width=0.850\columnwidth]{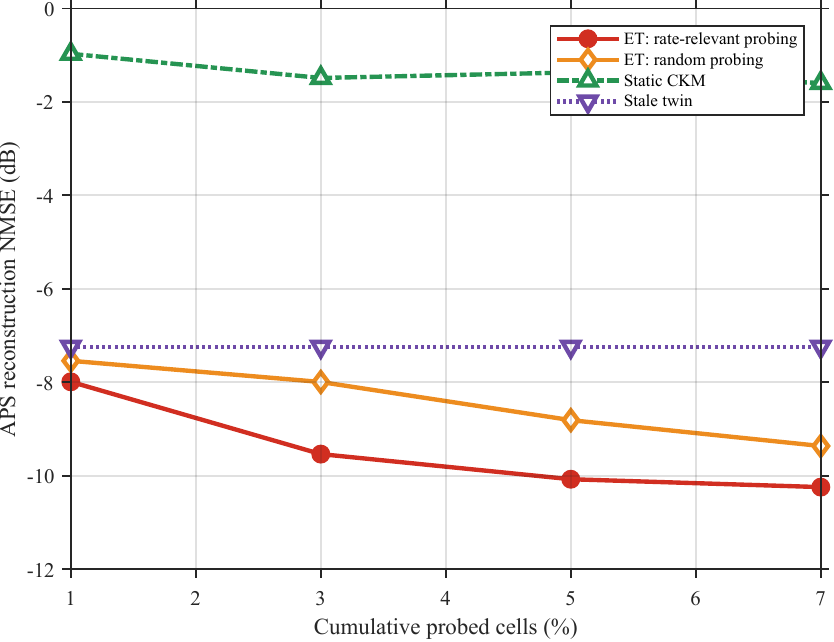}
\caption{20-update average closed-loop APS NMSE under rate-relevant and random probing.}
\label{fig:loop-nmse}
\end{figure}

Fig.~\ref{fig:loop-nmse} confirms that the rate gain accompanies a better state: at 7\%, rate-relevant probing gives $-10.24$ dB APS NMSE, versus $-9.36$, $-7.24$, and $-1.60$ dB for random ET, stale twin, and static CKM. Together, Figs.~\ref{fig:loop-rate} and~\ref{fig:loop-nmse} show the title's progression from uncertainty-reducing probes to improved covariances and sum rate.

\section{Conclusion}
This letter linked sparse probes to sum-rate maximization through ET beamforming. The ET retains an APS, updates its innovation, and queries user covariances; projected SCA designs the beams, while a rate-sensitivity bound selects the next probes. Results show the value of persistence under sparse evidence and the efficiency of rate-relevant probing. The loop thus specifies what wireless knowledge to retain, where to observe next, and how to use it for communication.


\begin{thebibliography}{99}
\ifdefined\WCLReview\else
\scriptsize
\setlength{\itemsep}{0pt}
\setlength{\parsep}{0pt}
\fi
\bibitem{romeroCartography}
D. Romero, S.-J. Kim, G. B. Giannakis, and R. L\'opez-Valcarce, ``Spectrum cartography: From sensor measurements to radio maps,'' \emph{IEEE Signal Process. Mag.}, vol. 32, no. 3, pp. 24--36, May 2015.

\bibitem{biRadioMap}
S. Bi, J. Lyu, Z. Ding, and R. Zhang, ``Engineering radio maps for wireless resource management,'' \emph{IEEE Wireless Commun.}, vol. 26, no. 2, pp. 133--141, Apr. 2019.

\bibitem{zengCKM}
Y. Zeng and X. Xu, ``Toward environment-aware 6G communications via channel knowledge map,'' \emph{IEEE Wireless Commun.}, vol. 28, no. 3, pp. 84--91, Jun. 2021.

\bibitem{ckmTutorial}
Y. Zeng \emph{et al.}, ``A tutorial on environment-aware communications via channel knowledge map for 6G,'' \emph{IEEE Commun. Surveys Tuts.}, vol. 26, no. 3, pp. 1478--1519, third quarter 2024.

\bibitem{saad6G}
W. Saad, M. Bennis, and M. Chen, ``A vision of 6G wireless systems: Applications, trends, technologies, and open research problems,'' \emph{IEEE Netw.}, vol. 34, no. 3, pp. 134--142, May/Jun. 2020.

\bibitem{wuFAS6G}
T. Wu \emph{et al.}, ``Fluid antenna systems enabling 6G: Principles, applications, and research directions,'' \emph{IEEE Wireless Commun.}, early access, 2025, doi: 10.1109/MWC.2025.3629597.

\bibitem{radioUNet}
R. Levie, C. Yapar, G. Kutyniok, and G. Caire, ``RadioUNet: Fast radio map estimation with convolutional neural networks,'' \emph{IEEE Trans. Wireless Commun.}, vol. 20, no. 6, pp. 4001--4015, Jun. 2021.

\bibitem{deepCompletion}
Y. Teganya and D. Romero, ``Deep completion autoencoders for radio map estimation,'' \emph{IEEE Trans. Wireless Commun.}, vol. 21, no. 3, pp. 1710--1724, Mar. 2022.

\bibitem{radioGAT}
X. Li \emph{et al.}, ``RadioGAT: A joint model-based and data-driven framework for multi-band radiomap reconstruction via graph attention networks,'' \emph{IEEE Trans. Wireless Commun.}, vol. 23, no. 11, pp. 17777--17792, Nov. 2024.

\bibitem{nguyenTwin}
H. X. Nguyen, R. Trestian, D. To, and M. Tatipamula, ``Digital twin for 5G and beyond,'' \emph{IEEE Commun. Mag.}, vol. 59, no. 2, pp. 10--15, Feb. 2021.

\bibitem{khanTwin}
L. U. Khan, W. Saad, D. Niyato, Z. Han, and C. S. Hong, ``Digital-twin-enabled 6G: Vision, architectural trends, and future directions,'' \emph{IEEE Commun. Mag.}, vol. 60, no. 1, pp. 74--80, Jan. 2022.

\bibitem{alkhateebTwin}
A. Alkhateeb, S. Jiang, and G. Charan, ``Real-time digital twins: Vision and research directions for 6G and beyond,'' \emph{IEEE Commun. Mag.}, vol. 61, no. 11, pp. 128--134, Nov. 2023.

\bibitem{wangDTC}
H. Wang \emph{et al.}, ``Digital twin channel for 6G: Concepts, architectures and potential applications,'' \emph{IEEE Commun. Mag.}, vol. 63, no. 3, pp. 24--30, Mar. 2025.

\bibitem{shumanGSP}
D. I. Shuman, S. K. Narang, P. Frossard, A. Ortega, and P. Vandergheynst, ``The emerging field of signal processing on graphs: Extending high-dimensional data analysis to networks and other irregular domains,'' \emph{IEEE Signal Process. Mag.}, vol. 30, no. 3, pp. 83--98, May 2013.

\bibitem{scutariSCA}
G. Scutari, F. Facchinei, L. Lampariello, and P. Song, ``Parallel and distributed methods for constrained nonconvex optimization---Part I: Theory,'' \emph{IEEE Trans. Signal Process.}, vol. 65, no. 8, pp. 1929--1944, Apr. 2017.

\bibitem{christensenWMMSE}
S. S. Christensen, R. Agarwal, E. de Carvalho, and J. M. Cioffi, ``Weighted sum-rate maximization using weighted MMSE for MIMO-BC beamforming design,'' \emph{IEEE Trans. Wireless Commun.}, vol. 7, no. 12, pp. 4792--4799, Dec. 2008.

\end{thebibliography}
\end{document}